# Drag coefficient for the air-sea exchange: foam impact in hurricane conditions.

Ephim Golbraikh,[1] and Yuri M. Shtemler,[2]

[1]*Department of Physics, Ben-Gurion University of the Negev, P.O. Box 653, Beer-Sheva 84105, Israel*

[2]*Department of Mechanical Engineering, Ben-Gurion University of the Negev, P.O. Box 653, Beer-Sheva 84105, Israel*

**Abstract.** A physical model is proposed for the estimation of the foam impact on the variation of the effective drag coefficient, $C_d$, with reference to the wind speed $U_{10}$ in stormy and hurricane conditions. In the present model $C_d$ is approximated by partitioning the sea surface into foam-covered and foam-free areas. Based on the available optical and radiometric measurements of the fractional foam coverage and the characteristic roughness of the sea-surface in the saturation limit of the foam coverage, the model yields the resulting dependence of $C_d$ vs $U_{10}$. This dependence is in fair agreement with that evaluated from field measurements of the vertical variation of the mean wind speed.

## 1. Introduction

Saturation/reduction of the effective drag coefficient, $C_d$, for the air-sea interface with wind speed rising up to hurricane (typhoon) conditions has been a focus of intensive experimental investigation over the last decade. Many field experiments (Powell et al., 2003; Black et al., 2007; Edson et al., 2007; Jarosz et al., 2007; Holthuijsen et al., 2012), laboratory (Donelan et al., 2004; Reul et al., 2008; Troitskaya et al., 2012), and theoretical studies (Bye and Jenkins 2006; Kudryavtsev and Makin, 2007; Bye and Wolff, 2008; Soloviev and Lukas, 2010 etc.) have been conducted to study variations of the sea-surface moment transfer and effective drag coefficient with wind speed in hurricane conditions. A reduction of the sea-surface drag coefficient in hurricane conditions instead of its monotonic growth with wind speed predicted by the Charnock relation commonly employed in moderate wind conditions (Charnock, 1955), has been found by Powell et al. (2003). As conjectured by Powell et al. (2003) and Holthuijsen et al. (2012), the foam cover increases due to wave breaking and forms a slip surface on the atmosphere - sea interface that leads to a saturation/reduction of the effective drag coefficient in hurricane conditions. Saturation in the drag coefficient growth has been observed in laboratory experiments by Donelan et al. (2004) who note that "one may expect a qualitatively different behavior in its properties than that suggested by observations in moderate wind conditions".

The principal role of the air–sea foam layer has been first suggested by Newell and Zakharov (1992). According to empirical data, foam formation is highly correlated with wind speed and sea gravity waves breaking (Stogryn, 1972; Monahan and O'Muircheartaigh, 1980; Monahan and Woolf, 1989; Reul and Chapron, 2003; Callaghan et al., 2007 etc.). The foam fractional coverage (foam fraction) monotonically increases with wind speed up to its saturation level (Holthuijsen et al., 2012). Properties of the near-surface water and the foam fractional coverage are changed when the wind speed increases (Camps et al., 2005; Boutin et al., 2012): foam salinity is dropping, while the main share of the foam coverage is formed by wind-aligned streaks, i.e. the "old" foam characterized by foam bubbles of larger sizes than those of white caps. In detail, wave breaking produces white caps together with wind-aligned streaks (Monahan and O'Muircheartaigh, 1980; Monahan and Woolf, 1989; Reul and Chapron, 2003; Callaghan et al., 2007; Holthuijsen et al., 2012). White caps production rises with wind speed and reaches its maximum when $U_{10}$ exceeds the storm strength $U_{10} \approx 25 ms^{-1}$ (see e.g. Powell et al., 2003; Anguelova and Webster, 2006; Anguelova and Peter, 2012; Holthuijsen et al., 2012). The coefficient of the foam fractional coverage up to white caps saturation stage is sufficiently small, $\alpha_f \sim 0.01 - 0.15$. In turn, wind-aligned streaks continue to grow with wind speed. As the wind reaches the hurricane strength ($U_{10} \approx 30 - 35 ms^{-1}$), wind-aligned streaks of foam bubbles combined with white caps cover the sea surface, and when $U_{10}$ reaches $\approx 40 ms^{-1}$, a foam layer covers the sea surface almost completely when the foam fraction approaches the saturation value (Reul and Chapron, 2003; Powell et al., 2003; El-Nimri et al., 2010; Holthuijsen et al., 2012).

Foam input into air-sea interaction in hurricane conditions was studied by Shtemler et al. (2010). A system with the foam has been modeled by a three-fluid system of the foam layer sandwiched between the atmosphere and the sea, by distributing foam spots homogeneously over the sea surface. They argued on physical grounds that the average roughness length for the foam -atmosphere interface should correlate with the characteristic size of the sea foam bubbles at hurricane wind speeds. Indeed, the characteristic size of the sea foam bubbles of the order of $0.1 - 2mm$ (Rayzer and Sharkov, 1980; Deane and Stokes, 2002; Leeuw and Leifer, 2002; Leifer et al., 2003; Soloviev and Lukas, 2006) well agrees with the experimental correlation for average aerodynamic roughness length ~ 0.1 – 2mm (Powell et al., 2003). Such a correlation between the aerodynamic and geometric roughness lengths at strong winds over the foamed sea surface distinguishes mobile systems from fixed beds. Namely, the aerodynamic roughness length of fixed beds significantly differs from the geometrical sizes of solid particles that constitute the beds (see a review by Dong et al., 2001 and references therein): for wind-blown sand surfaces Bagnold (1941) proposed a 1/30 law for the proportionality coefficient between the aerodynamic and geometric roughness. This law has been supported by Nikuradse's tests (Nikuradse, 1950) in pipes with inner walls artificially roughened by ideal sand grains of uniform radius. For non-ideal fixed beds, this coefficient depends on the wind speed and may vary significantly in a wide range of values. Fortunately, aerodynamic "roughness length has proven to be much more sensitive to the properties of ground surface" than other parameters, such as the surface drag coefficient and the effective surface momentum flux. For instance, for some fixed beds, the drag coefficient increases only $\sim 10^2$ times, while the aerodynamic roughness length increases $\sim 10^4$ times (Dong et al., 2001). They believe this is a reason why the aerodynamic roughness length has been widely used to characterize the aerodynamic properties of various fixed-bed surfaces. The difference between the aerodynamic roughness of fixed and mobile beds was also discussed by Dong et al. (2001), and they noted that mobile surfaces may adapt to the wind by changing roughness. For relatively weak winds blown over a mobile bed, such as water, the roughness length is well approximated by the well-known Charnock's (1955) equation. However, Charnock's equation predicts unrealistically high values of the effective drag coefficient for strong winds and should be substituted by a proper relation for the roughness of the sea surface foamed in hurricane conditions. In the absence of such relation, the sea surface roughness is conventionally determined by direct measurements of the wind speed up to storm and hurricane conditions at some distance above the sea surface and then extrapolated using the log-law model of the wind profile to the fictitious zero wind speed (Powell et al., 2003; Donelan et al. (2004); Edson et al. (2007); Black et al. (2007); Jarosz et al., 2007; Holthuijsen et al., 2012). In addition to the roughness length, $Z_0$, this procedure completely determines the effective values of the drag coefficient, $C_d$, and the surface friction velocity, $U_*$, vs. the wind speed at 10m reference height, $U_{10}$. This provides parameters of the logarithmic profile of the wind speed for further theoretical modeling of the atmosphere-sea interaction in hurricane conditions. For instance, Chernyavski et al. (2011) model the sea surface stability based on the effective aerodynamic roughness in the logarithmic wind profile instead of the effective aerodynamic roughness based on Charnock's formula. They also demonstrate that the wind stability model for hurricane conditions based on Charnock's formula underestimates by an order the growth rate of perturbations (the coefficient of the exponential growth of small perturbations of the air-sea interface induced by a logarithmic wind with time).

Such estimations of $Z_0$, $C_d$ and $U_*$ are based on the logarithmic law for wind profiles. At least, approximate validity of these assumptions in storm and hurricane conditions is the key point for such models (Tennekes, 1973). The applicability of Tennekes' (1973) theory to hurricane conditions is discussed by Smith and Montgomery (2014) (see also references therein). Remind that for the applicability of Tennekes' theory, the radial wind component should be negligibly small as compared with the tangential one, and they illustrate that this condition is approximately satisfied for storm and hurricane (typhoon) conditions with a relatively low error (lines 1 and 2 in Figure 4 in Smith and Montgomery, 2014). This is also supported by nearly vertical trajectories of the dropsondes observed during storm and typhoon stages of supertyphoon Jangmi (Sanger et al., 2014). Hence, for storm and hurricane conditions, relatively small deviations of the tangential wind component from the log-law may be expected. However, these conditions cannot be satisfied for supertyphoon stage (Sanger et al., 2014, see also lines 3 in Figure 4, Smith and Montgomery, 2014). Smith and Montgomery (2014) also demonstrate strong deviations from the log-law of the mean wind speed which results from averaging over several typhoons including a few supertyphoons (see Figure 7 in Smith and Montgomery, 2014). Powell et al. (2003) and Holthuijsen et al. (2012) argue that they obtain representative resulting mean wind profiles by averaging wind data sets obtained by grouping dropsondes with similar mean boundary layer wind speeds. They note that observations in all groups except the highest (supertyphoon) velocity follow the logarithmic profile. A detailed analysis of the two above-mentioned averaging procedures determining the mean wind speed is beyond the scope of this article. In the present study the existence of the logarithmic profile of mean wind speed is only assumed for storm and hurricane (typhoon) conditions, while the supertyphoon high winds are rejected from consideration.

The present study devoted to the estimation of the foam impact on the effective drag coefficient in storm and hurricane conditions is based on partitioning the sea surface into foam-covered and foam-free areas. For this purpose, the data of optical and radiometric measurements of the foam fractional coverage are used.

## 2. Physical model

The log-law model of the wind speed, $U$

$$U = (U_* / \varkappa) \ln(Z/Z_0),  \qquad (1)$$

where $\varkappa = 0.4$ is von Karman's constant; $Z$ [m] is the current height over the sea surface; $Z_0$ [m] is the sea aerodynamic surface roughness; $U_*$ [$ms^{-1}$] is the friction velocity. Together with the wind profile (1), the formula for the surface momentum flux $\tau = \rho U_*^2 = \rho C_{dL} U_L^2$ is commonly employed for the prediction of the drag coefficient, $C_{dL}$, variation with the neutral stability wind speed $U_L$ at a reference height $L$ [m] (c is the air density). It yields

$$C_{dL} = \frac{U_*^2}{U_L^2} = \left(\frac{\varkappa}{\ln(L/Z_0)}\right)^2. \qquad (2)$$

Conventionally, the drag prediction problem is solved by specifying the roughness length $Z_0$. Thus, for relatively weak winds, the roughness length is well approximated by the well-known formula (Charnock, 1955):

$$Z_0 = \sigma_{Ch} U_*^2 / g, \qquad (3)$$

where $g$ is an acceleration due to gravity, and $\sigma_{Ch}$ is a phenomenological constant. In the present paper, a standard value of the proportionality coefficient $y_{Ch} = 0.018$ has been adopted, which provides a better correspondence of the drag coefficient to the available experimental data at low winds (e.g. Large and Pond, 1981, Fairall et al., 2003, Edson et al., 2013).

At high wind speeds the aerodynamic roughness length of the sea surface totally covered by foam can be naturally related with the characteristic size of the foam bubbles, $R_b$ (by an analogy with fixed beds, see the discussion in Introduction):

$$Z_0 = \sigma_f R_b,$$

where both $\alpha_f$ and $R_b$ may vary with the wind speed. However, *in situ* measurements of $R_b$ are rather scarce, while $\sigma_f$ can be only estimated by the order of magnitude ($\sigma_f \sim 1$, Shtemler et al., 2010). Since these dependences are not well established, the present model assumes the roughness length of the foam-atmosphere interface at hurricane wind speeds given as a known physical constant:

$$Z_0 = const, \qquad (4)$$

where the constant will be specified below.

The friction force averaged over the surface, $S$, per unit area, which is caused by the viscous stress (surface momentum flux), $\tau$, is introduced as follows:

$$\tau = \frac{1}{S}\int_S \hat{\tau}dS' = \frac{1}{S}\left(\int_{S_w}\hat{\tau}_w dS' + \int_{S_f}\hat{\tau}_f dS'\right), \qquad (5)$$

where $S = S_w + S_f$, $S_w$ and $S_f$ are the total, water- and foam-covered areas, respectively, or alternatively, in terms of the mean viscous stresses

$$\tau = \frac{S-S_f}{S}\tau_w + \frac{S_f}{S}\tau_f.$$

This yields the following partitioning of the sea surface into foam-free and foam-covered areas (the partitioning rule) for the surface momentum flux $\tau = \rho U_*^2$ scaled by the air density:

$$U_*^2 = (1-\alpha_f)U_*^{(w)2} + \alpha_f U_*^{(f)2}, \qquad (6)$$

which reflects the energy conservation law and the additivity of energy losses per unit surface. Here $\alpha_f = S_f/S$ is the foam fractional coverage, $U_*^{(w)}$ and $U_*^{(f)}$ are friction velocities for the foam-free and foam-covered sea surfaces, respectively. Respectively, the first term in the right-hand side of formula (6) describes the surface momentum flux on the portion of the sea surface that is foam-free, while the second term gives it on the portion of the sea surface that is covered with foam patches. Both terms of Eq. (6) are completely determined by the observed foam coverage and the profile of the mean speed measured over the real waved and foamed sea-surface. This makes it evident that the input of all elements that constitute the total roughness is taken into account in (6). This type of area-weighted partitioning has been performed for a number of applications, e.g. for microwave emissivity (Stogryn, 1972) or temperature (Hosoda, 2010; Guimbard et al., 2012) of the foam-covered sea surface.

Applying relation (2) to each term in the left- and right-hand sides of formula (6), we have

$$U_*^2 = C_{dL}U_L^2, \quad U_*^{(w)2} = C_{dL}^{(w)}U_L^{(w)2}, \quad U_*^{(f)2} = C_{dL}^{(f)}U_L^{(f)2}.$$

Then formula (5) yields the following relation between drag coefficients at the reference length $L$:

$$C_{dL}U_L^2 = (1-\alpha_f)C_{dL}^{(w)}U_L^{(w)2} + \alpha_f C_{dL}^{(f)}U_L^{(f)2} \equiv U_*^2 = const. \qquad (7)$$

Since $U_*^2$ is a physical constant independent of the value of the reference height $L$, the present model adheres to the requirement of a constant momentum flux throughout the boundary layer. Indeed, equation (7) demonstrates that the model is self-adjusting for wind speeds measured at different heights to produce a constant flux layer, $C_{dL}U_L^2 \equiv const$ at any altitude located sufficiently far from the ocean surface, e.g. $C_{d5}U_5^2 = C_{d10}U_{10}^2$ for $L=5m$ and $L=10m$. Since the model inputs are adopted at 10 *m*,

$$C_{d10}U_{10}^2 = (1-\alpha_f)C_{d10}^{(w)}U_{10}^{(w)2} + \alpha_f C_{d10}^{(f)}U_{10}^{(f)2} \equiv U_*^2 = const. \qquad (8)$$

In addition, now we assume that the effective wind speed at any altitude $L$ located sufficiently far from the ocean surface has the same value as for the foam-free and totally foam-covered portions of the sea surface:

$$U_L = U_L^{(w)} = U_L^{(f)}. \qquad (9)$$

These relations have a meaning of the closure conditions for the model that relates the wind speeds at the reference height over different portions of the sea surface with the effective one. Then Eq. (8) can be rewritten in the form

$$C_{d10}\frac{U_{10}^2}{U_L^2}U_L^2 = (1-\alpha_{f10})C_{d10}^{(w)}\frac{U_{10}^{(w)2}}{U_L^{(w)2}}U_L^{(w)2} + \alpha_{f10}C_{d10}^{(f)}\frac{U_{10}^{(f)2}}{U_L^{(f)2}}U_L^{(f)2},$$

or, alternatively, with the help of log law (1) and Eqs. (9), as follows:

$$C_{d10} = (1-\alpha_f)C_{d10}^{(w)}\frac{A^{(w)}}{A} + \alpha_f C_{d10}^{(f)}\frac{A^{(f)}}{A}, \qquad (10)$$

where

$$A = \frac{\ln^2\left(\frac{10}{Z_0}\right)}{\ln^2\left(\frac{L}{Z_0}\right)}, \quad A^{(w)} = \frac{\ln^2\left(\frac{10}{Z_0^{(w)}}\right)}{\ln^2\left(\frac{L}{Z_0^{(w)}}\right)}, \quad A^{(f)} = \frac{\ln^2\left(\frac{10}{Z_0^{(f)}}\right)}{\ln^2\left(\frac{L}{Z_0^{(f)}}\right)}.$$

It is assumed here that the Charnock formula (3) for weak winds would also yield a fair approximation for high winds but for the presence of foam on the air-sea surface. As argued by Powell et al. (2003), the foam-covered sea-surface restricts the unbounded growth of the effective drag coefficient with wind speed. With this in mind, the first term in the right-hand side of Eq. (10) corresponds to a sea-surface portion completely free of foam, for which the Charnock formula (3) for the roughness length of the atmosphere-sea interface has been adopted: $Z_0^{(w)} = \sigma_{Ch}U_*^{(w)2}/g$, with $\sigma_{Ch} = 0.018$. Then, substituting $Z_0^{(w)}$ in formulas (2) yields implicit dependences of $C_{d10}^{(w)}$ and $U_*^{(w)}$ on $U_{10}^{(w)}$, where $Z_0^{(w)}/10$ satisfies the equation that follows from the Charnock formula

$$C_{d10}^{(w)} = \left(\frac{\varkappa}{\ln\left(10/Z_0^{(w)}\right)}\right)^2, \quad U_*^{(w)} = \sqrt{\frac{gZ_0^{(w)}}{\sigma_{Ch}}}, \quad \frac{Z_0^{(w)}}{10}\ln^2\left(\frac{Z_0^{(w)}}{10}\right) = \frac{\sigma_{Ch}\varkappa^2 U_{10}^{(w)2}}{g\,10}. \qquad (11)$$

The second term in Eq. (9) corresponds to the specific case of the portion of the sea surface entirely covered by foam. Similarly, $C_{d10}^{(f)}$ and $U_*^{(f)}$ are expressed by formulas (1) through the roughness length $Z_0^{(f)}$ of the sea-surface completely covered by foam:

$$C_{d10}^{(f)} = \left(\frac{\varkappa}{\ln(10/Z_0^{(f)})}\right)^2, \qquad U_*^{(f)} = \frac{\varkappa U_{10}^{(f)}}{\ln(10/Z_0^{(f)})}. \tag{12}$$

The present model contains six independent variables, $U_*^{(w)}$, $Z_0^{(w)}$, $U_*^{(f)}$, $Z_0^{(f)}$ and $U_*$, $Z_0$, that fully determine the logarithmic velocity profiles over the foam-free and totally foam-covered portions of the sea surface, as well as the effective logarithmic velocity profile averaged over alternating foam-free and foam-covered portions of the sea surface. Indeed, other dependent parameters of the logarithmic profiles such as $C_{d10}^{(w)}$, $U_L^{(w)}$, $C_{d10}^{(f)}$, $U_L^{(f)}$ and $C_{d10}$, $U_L$ can be expressed through $U_*^{(w)}$, $Z_0^{(w)}$, $U_*^{(f)}$, $Z_0^{(f)}$ and $U_*$, $Z_0$ using log law relations, such as (1) and (2). Five relations constitute the model, namely, the Charnock relation for the roughness length of the foam-free portion of the sea surface, $Z_0^{(w)}$, equation (4) for the roughness of the sea surface portion totally covered with foam, $Z_0^{(f)}$, along with two equations (9) and one equation (10) at the reference height $L$. These five equations for six independent variables can be reduced to a single equation, e.g. Eq. (10) that relates an effective drag coefficient $C_{d10}$ with $U_{10}$.

For further simplicity, the model can be tuned to the 10-m reference height, i.e.

$$U_{10} = U_{10}^{(w)} = U_{10}^{(f)}. \tag{13}$$

Then Eq. (7) is reduced to the following relation for the 10-m drag coefficients:

$$C_{d10} = (1 - \alpha_f)C_{d10}^{(w)} + \alpha_f C_{d10}^{(f)}. \tag{14}$$

Fractional foam coverage $\alpha_f$ is highly correlated to the wind speed $U_{10}$ (see Fig. 6c in Holthuijsen et al. (2012) and references therein). Although the total fractional foam coverage, which is a sum of whitecaps (Anguelova et al., 2006 and de Leeuw et al., 2011) and streaks, is of interest in the present study, the data of Holthuijsen et al. (2012) demonstrate that the input of white caps into $\alpha_f$ is negligibly small as compared with that for the streak coverage. So the observation data for $\alpha_f$ vs $U_{10}$ can be approximated as in Holthuijsen et al. (2012)

$$\alpha_f = \gamma \tanh[\alpha \exp(\beta U_{10})], \tag{15}$$

but with slightly varied values of $\alpha = 0.00255$, $\beta = 0.165$, $\gamma = 1.0$. These coefficients approximate experimental data as satisfactorily as the coefficients of Holthuijsen et al. (2012), but in addition they provide the total saturation at infinitely large wind speed. In Figure 1 the foam coverage $\alpha_f$ vs. $U_{10}$, following the model (15) is presented. As seen from (15), the fractional foam coverage is very close to unity beyond the wind speed $U_{10}$ of $40\ ms^{-1}$. While the fractional foam coverage is small when the wind is less than $25\ ms^{-1}$, it is significant as regards its influence on the ocean biogeochemistry (see, e.g., Vlahos and Monahan, 2009).

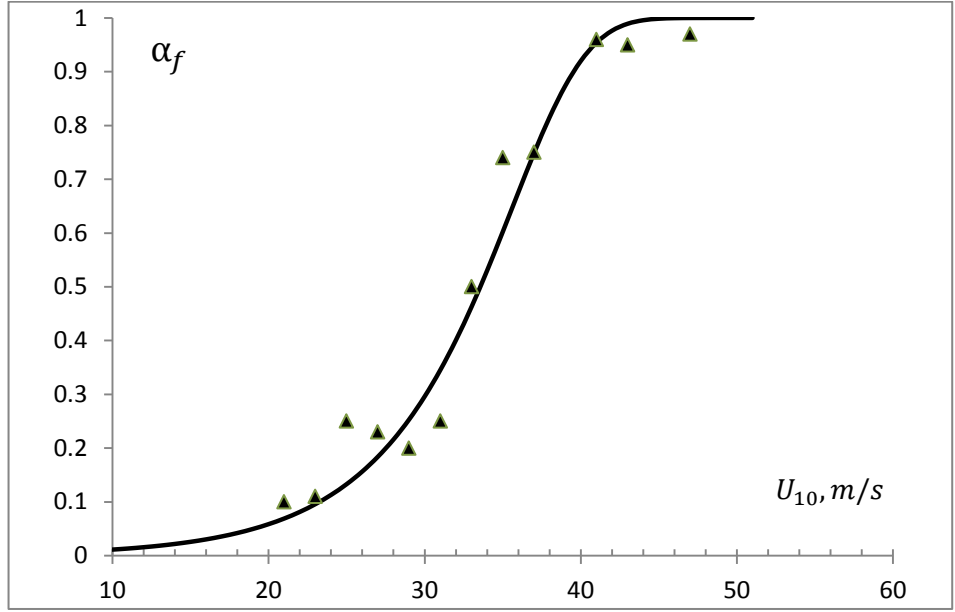

**Figure 1**. Foam coverage $\alpha_f$ vs. $U_{10}$.
Black triangles are adopted from observation data of Holthuijsen et al. (2012); solid line corresponds to the approximation (15).

Points in Figure 2 demonstrate the values of $C_{d10}$ evaluated from lab- (Donelan et al. (2004)) and field- (Powell et al. (2003); Edson et al. (2007); Black et al. (2007); Jarosz et al. (2007)) measurement data for the mean wind velocity in the range of $U_{10}$ ($0 < U_{10} < 50 ms^{-1}$). Meanwhile, Figures 3 and 4 depict the corresponding points for $Z_0$ and $U_*$ obtained using the above formulas (14) following from log-law for mean velocity.

To close the current model, the value $Z_0^{(f)}$ in (12) should be specified. As demonstrated in Figure 3, in hurricane conditions ($40\ ms^{-1} < U_{10} < 50\ ms^{-1}$) the field-measurement data for $Z_0$ are scattered within the range of $0.3\ mm$ to $1.6 mm$ around a constant mean value $< Z_0 > \approx 0.8\ mm$ (the lab- measurement data for $Z_0$ are dropped out from averaging that gives the mean value $< Z_0 > \approx 0.8\ mm$). Since the foam coverage converges to full saturation for wind speeds higher than $U_{10} \approx 40\ ms^{-1}$ (see Figure 6c in Holthuijsen et al., 2012), the above-mentioned constant mean value of the roughness can be naturally identified with the value

$Z_0^{(f)} = <Z_0>$.

Then, according to the current model, the effective drag coefficient, $C_{d10}$, is calculated vs. $U_{10}$ by the formulas (11)-(12) and (14)-(15). In a similar way, the dependence of the effective friction velocity, $U_*$, vs. $U_{10}$ may be obtained from relations (6), (11), (12) and (15). Substituting the already known $C_{d10}$ from (14) in Eq. (2), the effective roughness, $Z_0$, vs. $U_{10}$ is determined by the following relations:

$$Z_0 = 10 \exp\left(-\frac{\varkappa}{\sqrt{C_{d10}}}\right). \tag{16}$$

This yields the resulting dependences $C_{d10}$, $Z_0$ and $U_*$ vs. $U_{10}$ (Figures 2, 3 and 4) that are in a fair agreement with the data following from the field measurements of the vertical variation of the mean wind speed in the range from low to hurricane wind speeds. It is seen that the saturation of the resulting dependences of $C_{d10}$, $Z_0$ and $U_*$ vs. $U_{10}$ follows the saturation of the foam coverage $\alpha_f$ vs. $U_{10}$.

To evaluate the influence of the reference height $L$ on the results of modeling, we apply the method of successive iterations to relations (9) taking as the first approximation the values $Z_0$ that are obtained from the relation (14) for $C_{d10}$, which in turn follows from the condition (13). Our calculations show that the choice of the reference height $L$ has no practical effect on the value of the parameter $C_{d10}$. Thus, the dependence of $C_{d10}$ vs. $U_{10}$ obtained for the reference heights $L = 5$m or $L = 15$m deviate by less than one percent from that obtained for $L=10$. Since the measurement noise rather significantly exceeds the model deviations, it can be concluded that the reference height $L$, at which the condition $U_L = U_L^{(w)} = U_L^{(f)}$ is imposed, has no practical effect on the modeling results (with the only restriction that the reference length $L$ is located sufficiently far from the ocean surface as compared with the characteristic roughness length).

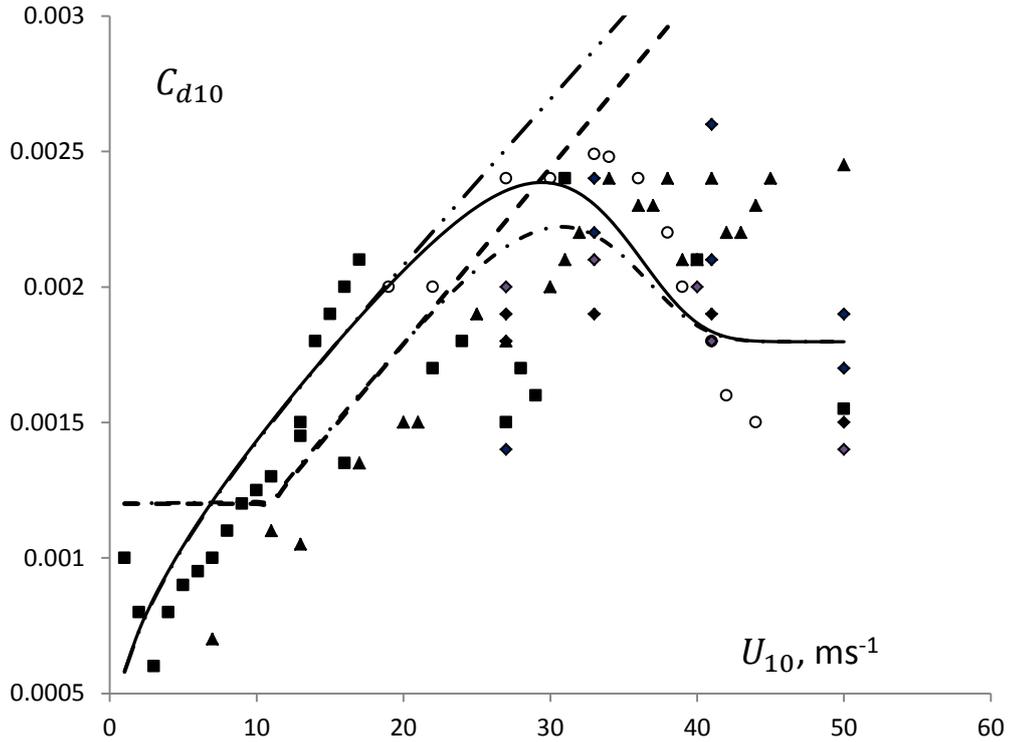

**Figure 2.** Effective drag coefficient $C_{d10}$ vs. $U_{10}$.

The line is $C_{d10}$ calculated with $C_{d10}^{(w)}$ adopted from Charnock's (1955) formula with $\sigma_{Ch} = 0.018$ and $Z_0^{(f)} = 0.8\ mm$, respectively. The heavy dashed-dot line is $C_{d10}$ with $C_{d10}^{(w)}$ adopted from Large and Pond (1981) and $Z_0^{(f)} = 0.8mm$. The dashed - double - dot and heavy-dashed lines are obtained with $C_{d10} = C_{d10}^{(w)}$, $C_{d10}^{(w)}$ being adopted from Charnock (1955) ($\sigma_{Ch} = 0.018$) and Large and Pond (1981), respectivelly. Triangles (Donelan et al., 2004), diamonds (Powell et al., 2003), squares (Edson et al., 2007 and Black et al., 2007), circles (Jarosz et al., 2007) are the points evaluated from the field measurements of the mean wind velocity.

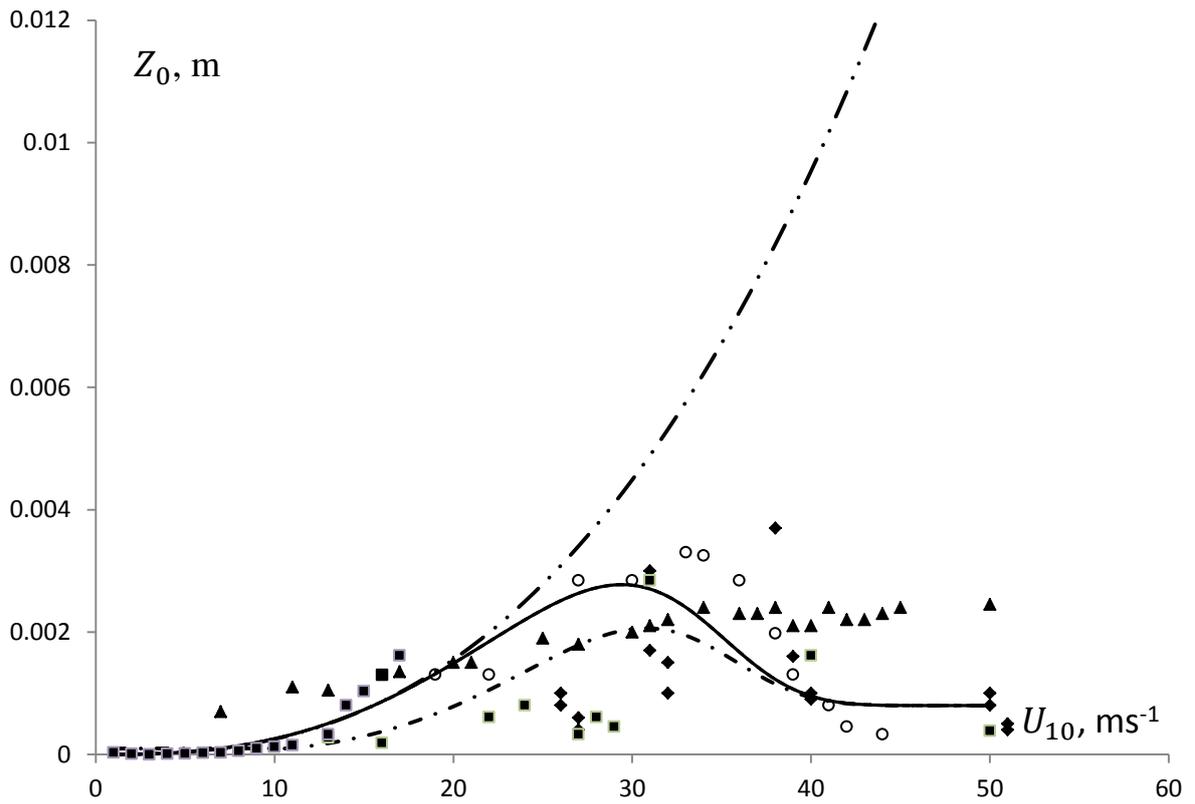

**Figure 3**. The effective roughness $Z_0$ vs. $U_{10}$ (notations as in Figure 2).

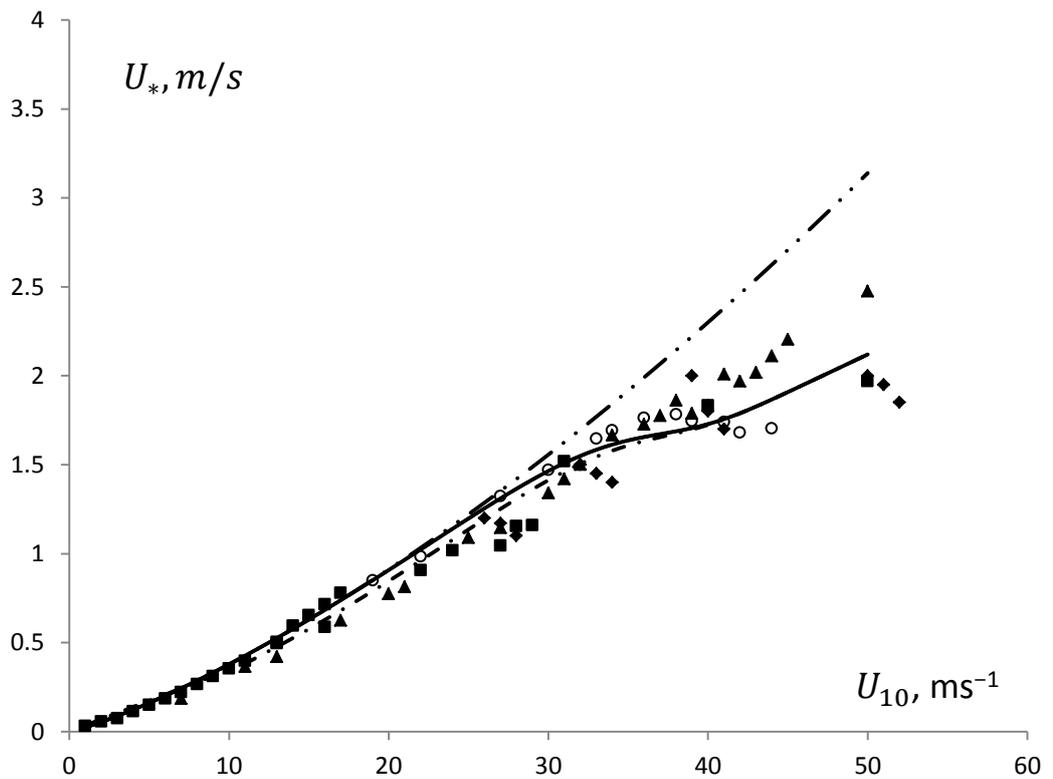

**Figure 4**. The effective friction velocity $U_*$ vs. $U_{10}$ (notations as in Figure 2).

To evaluate the influence of the reference height $L$ on the results of modeling, we apply the method of successive iterations to relations (10) taking as the first approximation the values $Z_0$ obtained from relation (14) for $C_{d10}$, which in turn follows from condition (13). Our calculations show that the choice of the reference height $L$ has no practical effect on the value of the parameter $C_{d10}$. Thus, the dependence of $C_{d10}$ vs. $U_{10}$ obtained for the reference heights $L = 5m$ or $L = 15m$ deviate by less than one percent from that obtained for $L = 10$. Since the measurement noise rather significantly exceeds the model deviations, it can be concluded that the reference height $L$ at which the condition $U_L = U_L^{(w)} = U_L^{(f)}$ is imposed has no practical effect on modeling results (with the only restriction that the reference length $L$ is located sufficiently far from the ocean surface as compared with the characteristic roughness length).

## 3. Summary and discussion

The present study is motivated by recent findings of saturation and even decrease in the drag coefficient (capping) in hurricane conditions, which is accompanied by the production of a foam layer on the ocean surface. The phenomenon of the drag-coefficient saturation/reduction is quite simply explained by wind slip conditions due to the separation of the atmosphere and sea by a foam layer produced by wave breaking at hurricane wind speeds. As it is difficult to expect at present a comprehensive numerical modeling of the drag coefficient saturation that is followed by wave breaking and foam production, there is no complete confidence and understanding of the saturation phenomenon. Since these results are obtained on the basis of field measurements of the vertical variation of the mean wind speed, the present study proposes an independent way to evaluate the drag coefficient based on partitioning the sea surface into foam-free and foam-covered surfaces.

The proposed model (formula (14)) treats the efficient air-sea drag coefficient, $C_{d10}$, as a sum of two weighted drag coefficients, $C_{d10}^{(w)}$ and $C_{d10}^{(f)}$, for the foam-free and foam-covered conditions. As accepted in the present model, each of the three drag coefficients, one on the left side and two on the right side of Eq. (14), should obey the log law, but at different interface conditions for the wind: real hurricane (i.e. alternating foam-free and foam-covered portions of the sea surface), foam-free and foam-covered sea-surface areas, respectively. This type of partitioning has been done for a number of applications (see references in Introduction), such as microwave emissivity or temperature of foam-covered sea surfaces, where the total emissivity or temperature is viewed as the area-weighted sum of the corresponding components at foam-covered and foam-free surfaces. Thus, for instance, the emissivities of the two classes of surface do not depend on each other, since they are determined only by the physical properties of the corresponding surfaces. In contrast, it is not obvious at all that the same separation is true for partitioning of an arbitrary characteristic. Thus, the partitioning of the sea surface into foam-free and foam-covered areas is evident for the square of friction velocity due to the additivity of energy losses, proven for the effective drag coefficient ((see relations (6) and (14) of the present model), while for the roughness, $Z_0$, this partitioning rule is not true at all. Indeed, according to equation (16) of our model, $Z_0$ varies in a nonlinear way with $\alpha_f$.

The model describes the variations of $C_{d10}$ with $U_{10}$ from very low to hurricane speed. The specific drag coefficient, $C_{d10}^{(w)}$, for the foam-free portions of the sea surface is modeled using the Charnock relation for roughness length determined by fitting the low wind data, while $C_{d10}^{(f)}$ for the foam-covered portions of the sea is modeled using the effective mean foam roughness in hurricane winds. The present approach is tuned to the reference height $L = 10\ m$, in the sense that the closure conditions of the model $U_L = U_L^{(w)} = U_L^{(f)}$ are imposed at this height, since all available experimental data are known at that altitude, and in this case the model is simplified. It is demonstrated that the choice of the reference height $L$ practically has no influence on the modeling results with the only restriction that $L$ is located sufficiently far from the ocean surface, so that $L$ significantly exceeds the roughness values.

The current model is based on the available optical and radiometric measurements of the fractional foam coverage, $\alpha_f$, combined with direct wind speed measurements in the hurricane conditions which provide the mean roughness of the sea surface totally covered with foam, $Z_0^{(f)}$. In particular, the present model yields $C_{d10}$ vs. $U_{10}$ in fair agreement with that evaluated from field measurements of the vertical variation of mean wind speed in the range $U_{10}$ from low to hurricane speeds (Powell et al., 2003; Edson et al., 2007; Black et al., 2007; Jarosz et al., 2007; Holthuijsen et al., 2012). The present approach opens opportunities for drag coefficient modeling in hurricane conditions using optical and radiometric measurements which have been intensively developed during the last two decades (Amarin et al. 2012, and references therein). For further improvement of the proposed model, $\alpha_f$ and $Z_0^{(f)}$ should be evaluated over a wider range of measurements and with the account for the influence of other physical parameters (such as temperature, salinity etc., Holthuijsen et al., 2012).

## References

Amarin RA, Jones WL, El-Nimri SF, Johnson JW, Ruf CS, Miller TL, Uhlhorn E (2012) Hurricane Wind Speed Measurements in Rainy Conditions Using the Airborne Hurricane Imaging Radiometer (HIRAD). IEEE T. Geosc. Remote 50(1):180-192

Anguelova MD, Webster F (2006), White cap coverage from satellite measurements: A first step toward modeling the variability of oceanic white caps, J. Geophys. Res. 111:C03017

Anguelova MD, Peter WG (2012) Dielectric and Radiative Properties of Sea Foam at Microwave Frequencies: Conceptual Understanding of Foam Emissivity. Remote Sens 4:1162-1189

Bagnold RA, *The Physics of Blown Sand and Desert Dunes,* Methuen, New York, 1941, 265 pp.

Black PG, D'Asaro EA, Drennan WM, French JR, Niiler PP, Sanford TB, Terrill EJ, Walsh EJ, Zhang JA (2007) Air-sea exchange in hurricanes: Synthesis of observations from the Coupled Boundary Layer Air-Sea Transfer experiment. B Am Meteor Soc 88(3): 357–374

Bye JAT, Jenkins AD (2006) Drag coefficient reduction at very high wind speeds. J Geophys Res 111:C03024

Boutin J, Martin N, Yin X, Font J, Reul N, Spurgeon P (2012) First assessment of SMOS data over open ocean: Part II sea surface salinity. IEEE Trans Geosc Remote Sensing. 50(5): 1662-1675

Bortkovskii R, *Air-Sea Exchange of Heat and Moisture During Storms*, Springer, New York, 1987, 193 pp.

Bye JAT, Wolff J-O (2008), Charnock dynamics: A model for the velocity structure in the wave boundary layer of the air-sea interface. Ocean Dyn 58:31–42

Callaghan A, De Leeuw G, Cohen L (2007) Observations of Oceanic Whitecap Coverage in the North Atlantic during Gale Force Winds. Nucl Atm Aerosols 9:1088-1092

Camps A, Vall-Ilossera M, Villarino R, Reul N, Chapron B, Corbella I, Duffo N, Torres F, Miranda Jj, Sabia R, Monerris A, Rodriguez R (2005) The Emissivity of Foam-Covered

Water Surface at L-Band: Theoretical Modeling and Experimental Results From the Frog 2003 Field Experiment. IEEE Trans Geosc Remote Sensing. 43(5): 925-937

Charnock H (1955) Wind stress on a water surface. Q J R Meteorol Soc 81: 639-640

Chernyavski VM, Shtemler YM, Golbraikh E, Mond M (2011) Generation of intermediately long sea waves by weakly sheared winds Phys Fluids 23: 016604-1-5

Deane GB, Stokes MD (2002) Scale dependence of bubble creation mechanisms in breaking waves. Nature 418:839-834

Donelan MA, Haus BK, Reul N, Plant W, Stiassnie M, Graber H, Brown O, Saltzman E (2004) On the limiting aerodynamic roughness of the ocean in very strong winds. Geophys Res Lett 31: L18306

Dong Z, Wang W, Zhao, Liu L, Liu X (2001) Aerodynamic roughness of fixed sandy beds. J Geoph Res VOL. 106(6): 11,001-11,011

Edson J, Crawford T, Crescenti J, Farrar T, French J, Frew N, Gerbi G, Helmis C, Hristov T, Khelif D, Jessup A, Jonsson H, Li M, Mahrt L, McGillis W, Plueddemann A, Shen L, Skyllingstad E, Stanton T, Sullivan P, Sun J, Trowbridge J, Vickers D, Wang S, Wang Q, Weller R, Wilkin J, Yu D, Zappa C (2007) The Coupled Boundary Layers and Air-Sea Transfer Experiment in Low Winds (CBLAST-LOW). Bull Amer Meteorol Soc 88(3): 341-356

Edson JB, Jampana V, Weller RA, Bigorre SP, Plueddemann AJ, Fairall CW, Miller SD, Mahrt L, Vickers D, Hersbach H (2013) On the Exchange of Momentum over the Open Ocean. J. Phys. Oceanogr. 43:1589–1610

El-Nimri SF, Jones WL, Uhlhorn E, Ruf C, Johnson J, Black P (2010) An Improved C-Band Ocean Surface Emissivity Model at Hurricane-Force Wind Speeds Over a Wide Range of Earth Incidence Angles. IEEE Geosc. Remote 7(4):641-645

Fairall CW, Bradley EF, Hare JE, Grachev AA, Edson JB (2003) Bulk Parameterization of Air–Sea Fluxes. Updates and Verification for the COARE Algorithm. J. Climate 16:571-591

Guimbard S, Gourrion J, Portabella M, Turiel A, Gabarró C, Font J (2012) SMOS Semi-Empirical Ocean Forward Model Adjustment. IEEE Trans. Geosc. Remote Sensing 50(5): 1676-1687

Holthuijsen LH, Powell MD, Pietrzak JD (2012) Wind and waves in extreme hurricanes. J. Geoph. Res. 117:C09003-1-15

Jarosz E, Mitchell DA, Wang DW, Teague WJ (2007) Bottomup determination of air-sea momentum exchange under a major tropical cyclone. Science 315:1707–1709

Kudryavtsev V.N., Makin V.N. (2007) Aerodynamic roughness of the sea surface at high winds. Boundary Layer Meteorol. 125: 289–303

Large WG, Pond S (1981) Open Ocean Flux Measurements in Moderate to strong Winds, J. Phys. Ocean 11: 324-336

de Leeuw G and Leifer I. (2002) Bubbles Outside the Plume During the LUMINY Wind-Wave. Gas Transfer at Water Surfaces, Geophysical Monograph 127: 295-301

de Leeuw, G., E. L Andreas, M. D. Anguelova, C. W. Fairall, E. R. Lewis, C. O'Dowd, M. Schulz, and S. E. Schwartz, 2011: Production flux of sea-spray aerosol. *Rev. Geophys.*, **49** (RG2001), doi:10.1029/2010RG000349.

Leifer I, De Leeuw G and Cohen LH (2003) Optical Measurement of Bubbles: System Design and Application. J. of Atmosp. Oceanic Techn. 20(9):1317-1332

Monahan E.C., O'Muircheartaigh I. (1980) Optimal Power-Law Description of Oceanic Whitecap Coverage Dependence on Wind Speed. J. Phys. Oceanogr. 10, 2094-2099

Monahan E., Woolf D.K. (1989) Comments on Variations of Whitecap Coverage with Wind Stress and Water Temperature. . J. Phys. Oceanogr. 19:706-709

Newell AC, Zakharov VE (1992) Rough ocean foam. Phys. Rev. Lett. 69:1149–1151

Nikuradse J, Laws of flow in rough pipes, *Tech. Memo. 1292,* Natl. Advis. Comm. on Aeronaut., Washington, D. C., 1950.

Powell MD, Vickery PJ, Reinhold TA (2003) Reduced drag coefficient for high wind speeds in tropical cyclones. Nature 422:279–283

Rayzer VY, Sharkov YA (1980) On the dispersed structure of the sea foam. Izvestia RAN: Atm. Ocean Phys. 16(7): 548-550

Reul N, Chapron B (2003) A Model of Sea-Foam Thickness Distribution for Passive Microwave Remote Sensing Applications. J. Geophys. Res. 108:19.4 - 19.14

Reul N, Branger H, Giovanangeli, J-P (2008) Air flow structure over short-gravity breaking waves. Boundary-Layer Meteorol. 126:477–505

Sanger NT, Montgomery MT, Smith RK, Bell MM (2014) An Observational Study of Tropical Cyclone Spinup in Supertyphoon Jangmi (2008) from 24 to 27 September. Mont. Weather Rev. 142: 3-28

Shtemler YM, Golbraikh E, Mond M (2010) Wind–wave stabilization by a foam layer between the atmosphere and the ocean. Dyn. Atmosph. and Oceans 50: 1–15

Smith RK, Montgomery MT (2014) On the existence of the logarithmic surface layer in the inner core of hurricanes. Q. J. R. Meteorol. Soc. 140: 72–81

Soloviev A and Lukas R. The Near-Surface Layer of the Ocean: Structure, Dynamics and Applications. Springe, Dordrecht, Netherlands, 2006, 572 pp.

Soloviev A, Lukas R. (2010) Effects of bubbles and spray on air-sea exchange in hurricane conditions. Boundary Layer Meteorol. 136:365–376

Stogryn A (1972) The Emissivity of Sea Foam at Microwave Frequencies. J. Geophys. Res. 77: 1658-1666

Tennekes H 1973. The logarithmic wind profile. *J. Atmos. Sci.* 30:234–238.

Stramska M, and Petelski T (2003) Observations of oceanic whitecaps in the north polar waters of the Atlantic. J. Geophys. Res., 108(C3): 308631-1-10

Troitskaya YI, Sergeev DA, Kandaurov AA, Baidakov GA, Vdovin MA, Kazakov VI (2012) Laboratory and theoretical modeling of air-sea momentum transfer under severe wind conditions. J. Geophys. Res. 117:C00J21-1-13

Vlahos P, Monahan EC (2009) A generalized model for the air-sea transfer of dimethyl sulfide at high wind speeds. Geophys. Res. Lett. 36: L21605

Wu J (1988) Variations of whitecap coverage with wind stress and water temperature. J. Phys. Oceanogr. 18:1448–1453.

Zhang JA, Rogers RF, Nolan DS, Frank M Jr. (2011) On the Characteristic Height Scales of the Hurricane Boundary Layer. Monh. Weather. Rev. 139: 2523-2535